\documentclass[aps,prl,twocolumn,superscriptaddress,showpacs]{revtex4}
\usepackage{epsfig,pstricks,dcolumn}

\begin{document}
%\preprint{}

%Title of paper
\title{Symmetry energy of dilute warm nuclear matter }
\author{J.~B.~Natowitz}
%\email{natowitz@comp.tamu.edu}
\affiliation{Cyclotron Institute, Texas A \& M University, College Station, 
Texas 77843-3366, USA}
\author{G.~R\"{o}pke}
%\email{gerd.roepke@uni-rostock.de}
\affiliation{Institut f\"{u}r Physik, Universit\"{a}t Rostock,
Universit\"{a}tsplatz 3, D-18055 Rostock, Germany}
\author{S.~Typel}
%\email{s.typel@gsi.de}
\affiliation{Excellence Cluster Universe, 
Technische Universit\"{a}t M\"{u}nchen, Boltzmannstra\ss{}e 2, D-85748
Garching, Germany}
\affiliation{GSI Helmholtzzentrum f\"{u}r Schwerionenforschung GmbH, 
Theorie, Planckstra\ss{}e 1, D-64291 Darmstadt, Germany}
\author{D.~Blaschke}
%\email{blaschke@ift.uni.wroc.pl}
\affiliation{Instytut Fizyki Teoretycznej, Uniwersytet Wroc{\l}awski,
pl. M. Borna 9, 50-204 Wroc{\l}aw, Poland}
\affiliation{Bogoliubov Laboratory for Theoretical Physics, JINR Dubna,
Joliot-Curie str. 6,  141980 Dubna, Russia
}
\author{A.~Bonasera}
%\email{}
\affiliation{Cyclotron Institute, Texas A \& M University, College Station, 
Texas 77843-3366, USA}
% and Laboratori Nazionali del Sud-INFN, v. S.Sofia 64 95123 Catania, Italy.}
\affiliation{Laboratori Nazionali del Sud-INFN, v. S. Sofia 64, 95123 Catania, 
Italy.}
\author{K.~Hagel}
%\email{}
\affiliation{Cyclotron Institute, Texas A \& M University, College Station, 
Texas 77843-3366, USA}   
\author{T.~Kl\"{a}hn}
%\email{thomas.klaehn@googlemail.com}
\affiliation{Instytut Fizyki Teoretycznej, Uniwersytet Wroc{\l}awski,
pl. M. Borna 9, 50-204 Wroc{\l}aw, Poland}
\affiliation{Theory Group, Physics Division, Building 203,
Argonne National Laboratory,
9700 South Cass Avenue, Argonne, IL 60439, USA}
\author{S.~Kowalski}
%\email{}
\affiliation{Cyclotron Institute, Texas A \& M University, College Station, 
Texas 77843-3366, USA}
\author{L.~Qin}
%\email{}
\affiliation{Cyclotron Institute, Texas A \& M University, College Station, 
Texas 77843-3366, USA}
\author{S.~Shlomo}
%\email{}
\affiliation{Cyclotron Institute, Texas A \& M University, College Station, 
Texas 77843-3366, USA}
\author{R.~Wada}
%\email{}
\affiliation{Cyclotron Institute, Texas A \& M University, College Station, 
Texas 77843-3366, USA}
\author{H.~H.~Wolter}
%\email{hermann.wolter@physik.uni-muenchen.de}
\affiliation{Fakult\"at f\"ur Physik, Universit\"{a}t M\"{u}nchen,
Am Coulombwall 1, D-85748 Garching, Germany}

\begin{abstract}
The symmetry energy of nuclear matter is a fundamental ingredient in
the investigation of exotic nuclei, heavy-ion collisions and
astrophysical phenomena.
New data from heavy-ion collisions can be used to extract the free symmetry 
energy and the internal symmetry energy at subsaturation 
densities and temperatures below 10~MeV.
Conventional theoretical calculations of
the symmetry energy based on mean-field approaches 
fail to give the correct low-temperature, low-density limit that is governed 
by correlations, in particular by the appearance of bound states.
A recently developed quantum statistical (QS) approach 
that takes the formation of clusters into account predicts 
symmetry energies
that are in very good agreement with the experimental data. 
A consistent description of the symmetry energy is given that joins the 
correct low-density limit 
with quasiparticle approaches valid near the saturation density.
\end{abstract}

\date{\today}

\maketitle

The symmetry energy in the nuclear equation of state governs
phenomena from the structure of exotic nuclei to astrophysical processes. 
The structure and the composition of neutron stars
depend crucially on
the density dependence of the symmetry energy
\cite{Lattimer:2006xb}. 
The symmetry energy characterizes
the dependence of the nuclear binding energy on the 
asymmetry $\delta=(N-Z)/A$
with $Z$ and $N$
the proton and neutron numbers, and $A=N+Z$.
As a general representation of the symmetry energy coefficient we use the
definition
 \begin{equation}
\label{eq:esym_def}
  E_{\rm sym}(n,T) = \frac{E(n,1,T)
  + E(n,-1,T)}{2} - E(n,0,T) ,
\end{equation}
where $E(n,\delta,T)$ is the energy per nucleon  
of nuclear matter with density $n$, asymmetry $\delta$, and temperature $T$.
At low density the symmetry energy changes mainly because 
additional binding is gained in symmetric matter due to 
formation of 
clusters and pasta structures \cite{Watanabe:2009vi}.

Our empirical knowledge of the symmetry energy near the saturation
density, $n_{0}$, is based primarily on the binding energies of nuclei.
The Bethe-Weizs\"{a}cker mass formula leads to values of about 
$E_{\rm sym}(n_0,0)=28-34$ MeV 
for the symmetry energy at zero temperature 
and saturation density $n_{0} \approx 0.16$ fm$^{-3}$, if 
surface asymmetry effects are properly taken into account \cite{Dan03}.

\begin{table*} [!t]
\caption{\label{tab:01}
Temperatures, densities, free and internal symmetry energies 
for different values of the surface velocity as derived from heavy-ion 
collisions 
(cols. 2-6), from the QS approach 
(cols. 7-8) and selfconsistently with clusters (cols. 9-12), see text.}
\begin{ruledtabular}
\begin{tabular}{|c|ccccccc|cccc|}
   $V_{\rm surf}$
 & $T$
 & $n$
 & $F_{\rm sym} $
 & $S_{\rm sym}^{\rm NSE} $
 & $E_{\rm sym} $
 & $F_{\rm sym}^{\rm QS} $
 & $E_{\rm sym}^{\rm QS} $
 & $T^{\rm sc}$
 & $n^{\rm sc}$
 & $F_{\rm sym}^{\rm sc}$
 & $E_{\rm sym}^{\rm sc}$
 \\
  (cm/ns)
  & (MeV)
  & (fm$^{-3}$)
  &(MeV)
  & (${\rm k}_{\rm B}$)
  & (MeV)
  &(MeV)
  & (MeV)
  &(MeV)
  &(fm$^{-3}$)
  & (MeV)
  &(MeV)
  \\
\hline
0.75 &3.31 &0.00206  &  5.64 & 0.5513 & 7.465 & 6.607 & 8.011 & 3.26 & 0.00493 & 9.211& 9.666\\
1.25 &3.32 &0.00165  &  6.07 & 0.5923 & 8.036 & 6.087 & 7.502 & 3.45 & 0.00511 & 9.295 &
9.647\\
1.75 &3.61 &0.00234  &  6.63 & 0.4137 & 8.124 & 6.877 & 7.896 & 3.54 & 0.00510 & 9.284 &
9.612\\
2.25 &4.15 &0.00378  &  7.81 & 0.1557 & 8.456 & 8.184 & 8.305 & 3.66 & 0.00495 & 9.193 &
9.524\\
2.75 &4.71 &0.00468  &  8.28 & -0.0162 & 8.204 & 8.967 & 8.321 & 4.02 & 0.00510 & 9.274 &
9.386\\
3.25 &5.27 &0.00489  &  9.30 & -0.1358 & 8.584 & 9.395 & 7.785 & 4.65 & 0.00574 & 9.683 &
9.227\\
3.75 &6.24 &0.00549  &10.69  & -0.2936 & 8.858 & 10.729 & 7.623 & 5.75 & 0.00684 & 10.487 &
8.978\\
4.25 &7.54 &0.00636 &11.83  & -0.4197 & 8.665 & 11.397 & 7.807 & 7.46 & 0.00866 & 11.982 &
8.964\\
\end{tabular}
\end{ruledtabular}
\end{table*}

In contrast to the value of 
$E_{\rm sym}(n_0,0)$, 
the variation of the symmetry energy 
with density and temperature is intensely debated.
Many theoretical investigations have been performed to estimate the 
behavior of the symmetry energy as a function of $n$ and $T$. 
A recent review is given by Li et al.\ \cite{Li:2008gp}, see also 
\cite{Fuchs:2005yn,Klahn:2006ir}. 
Typically, quasiparticle approaches such as the 
Skyrme Hartree-Fock and relativistic mean 
field (RMF) models or Dirac-Brueckner Hartree-Fock (DBHF) 
calculations are used. In such calculations the symmetry 
energy tends to zero in the low-density limit
for uniform matter. 
However, in accordance with the mass action law, 
cluster formation dominates the structure of low-density symmetric matter at 
low temperatures. 
Therefore, the symmetry energy in this low-temperature limit has to be equal 
to the binding energy per nucleon associated with the strong interaction 
of the most bound nuclear cluster.
A single-nucleon quasiparticle approach cannot account for such structures.
The correct low-density limit can be recovered only if 
the formation of clusters is properly taken into account,
as has previously been shown in Ref. \cite{Horowitz:2005nd}
in the context of a virial expansion valid at very low densities,
and in Ref.\ \cite{Typel:2009sy}.

In this letter we employ a quantum statistical (QS) approach which
includes cluster correlations in the medium. It
interpolates between the exact low-density limit and the 
very successful RMF approaches near the saturation density.
We show that this picture is 
in agreement with recent experimental findings on 
$E_{\rm sym}$ at very low densities.

Suitable approaches to account for cluster formation 
are the nuclear statistical equilibrium (NSE)
model \cite{Bondorf:1995ua}, 
cluster-virial expansions 
\cite{Horowitz:2005nd}, and generalized Beth-Uhlenbeck approaches 
\cite{Schmidt:1990}.
A thermal Green
function approach that allows a generalization of the NSE model by introducing
a quasiparticle description also for the bound states was already formulated 
some decades ago  \cite{Ropke:1983}, but only recently
 analyzed
with respect to the consequences for nuclear matter 
\cite{Ropke:2008qk}. 
In this QS approach the 
cluster correlations are described in a generalized Beth-Uhlenbeck expansion.
The advantage of this method is that the medium modifications of the 
clusters at finite density are taken into account.
In Ref.~\cite{Typel:2009sy} the thermodynamic properties of nuclear
matter were derived using this approach. 
The formulation of Ref.~\cite{Typel:2009sy} is valid 
in the density and temperature range
where the formation of light clusters with 
$A \leq 4$ dominates and heavier clusters are not yet important.
The method requires a sufficiently accurate model for the 
quasiparticle properties, for which we employ 
a RMF model with density dependent couplings
\cite{Typ2005}
which gives a good description both of nuclear matter around 
normal density and of ground state properties of nuclei across the nuclear 
chart.
In order to extend the applicability of this RMF model to 
very low densities, it has been generalized 
in Ref.\ \cite{Typel:2009sy} to account also for  
cluster formation and dissolution. 

We note that at very low densities and temperatures
below $T\simeq1$~MeV new phases may occur.
In fact, the formation 
of a solid phase using Overhauser orbitals 
including a triple point~\cite{Jaqaman88} or 
Bose-Einstein condensation~\cite{Fuku04}  have been suggested. 
However, in this letter 
we are concerned with  experimental data which probe nuclear matter 
at considerably higher temperatures. 
The low-$T$ behavior is an interesting issue for future studies.  

In the following we focus on finite temperatures and 
on the sub-saturation region $n < n_{0}$. 
Experimental information is derived from heavy-ion 
collisions of charge asymmetric nuclei, where transient states of different 
density can be reached, depending on the incident energy and the centrality of 
the collision. 
In the Fermi energy domain  symmetry energy effects have 
been investigated using judiciously 
chosen observables 
\cite{Baran:2004ih,Li:2008gp,Tsang:2008fd,Wuenschel:2009,Sfienti09}. 

Recently, the experimental determination of the 
symmetry energy at very low densities 
produced in heavy ion collisions of $^{64}$Zn on $^{92}$Mo and $^{197}$Au 
at 35~MeV per nucleon has 
been reported \cite{Kowalski:2006ju}.
Results of this study are given in the first four columns of Tab.\ 
\ref{tab:01}. 
Note that as a result of an energy recalibration and 
reevaluation of the particle yields in different velocity bins these 
values are slightly different than those reported in Ref.  
\cite{Kowalski:2006ju}. 
The surface velocity $v_{\rm surf}$, i.e.\ the velocity before the 
final Coulomb acceleration, was 
used as a measure of the time when the particles leave the source 
under different conditions of density and temperature. 
Only values of $v_{\rm surf} < 4.5$~cm/ns are included here, since 
the system does not reach equilibrium 
for higher $v_{\rm surf}$, see Tab.~I of Ref.~\cite{Kowalski:2006ju}. 
The yields of the light clusters $A \le 4$
were determined as a function of  
$v_{\rm surf}$. 
Temperatures were determined with
the Albergo method \cite{Albergo85}
using a H-He thermometer based on the double yield 
ratio of deuterons, 
tritons, $^3$He and $^4$He, and are given in Tab. \ref{tab:01}
as the average for the two reactions. 

The  free neutron yield is obtained from the free proton yield and the yield 
ratio of ${}^{3}$H to $^3$He. To determine
the asymmetry parameter of the sources
the total
proton and neutron yields including those bound in clusters are
used.
The proton chemical potential is derived from the yield 
ratio of ${}^{3}$H to  $^4$He. 
The corresponding free proton and free neutron densities are calculated, and 
the total nucleon density is obtained by accounting also for the bound 
nucleons according to their respective yields \cite{Kowalski:2006ju}.
The total nucleon densities are of the order 
of 1/100th to 1/20th of saturation density, as seen in 
Tab.~\ref{tab:01}.

An isoscaling analysis \cite{MBTsang}
has been employed (as a function of $v_{\rm surf}$) to 
determine the free symmetry energy $F_{\rm sym}$ via the expression 
$\alpha=4F_{\rm sym}\, \Delta(Z/A)^2/T$.
Here $\alpha$ is the isoscaling coefficient determined from yield ratios of 
$Z = 1$ ejectiles of the two reactions and $ \Delta(Z/A)^2$ is the difference
of the squared asymmetries
of the sources in the two reactions.
With $ \Delta(Z/A)^2$ and the temperature determined as above, the 
free symmetry energy is extracted. 

From the free symmetry energy derived in this way from the 
measured yields, the internal 
symmetry energy can be calculated if the symmetry entropy is known. 
The values of the symmetry entropy $S^{\rm NSE}_{\rm sym}$
for given parameters of temperature and 
density within the NSE model are shown in  Tab.\ \ref{tab:01}, column 5. 
They are calculated with the equivalent expression of Eq. (1) as the
difference between the entropies of pure proton or neutron and symmetric 
nuclear matter. 
In contrast to the mixing entropy that leads to a larger entropy for 
uncorrelated symmetric matter in 
comparison with pure neutron matter, the formation of correlations, in 
particular clusters, will reduce the
entropy in symmetric matter, see also Fig.~9 of Ref.~\cite{Typel:2009sy}.
For parameter values for which the yields of free nucleons in symmetric matter 
are small, the symmetry entropy may become positive, as seen in  
Tab.~\ref{tab:01} for low temperatures. 
The fraction of nucleons bound in clusters can decrease, e.g. due to 
increasing temperature or the dissolution of bound states at high densities 
due to the Pauli blocking. 
Then, the symmetric matter recovers its larger entropy so that the symmetry 
entropy becomes negative, as seen  in  Tab.~\ref{tab:01} also in the QS and 
self-consistent (sc, see below) calculations.

The results obtained in this way for the internal 
symmetry energy $E_{\rm sym}=F_{\rm sym} +T S^{\rm NSE}_{\rm sym}$ are
shown in Tab.~\ref{tab:01}, column 6.
We note that in Ref.~\cite{Kowalski:2006ju} the 
symmetry entropy was estimated using results of the 
virial expansion of Ref. \cite{Horowitz:2005nd} leading to
different internal symmetry energies. 
However, this approximation is unreliable at the 
densities considered here.

\begin{figure}
\epsfig{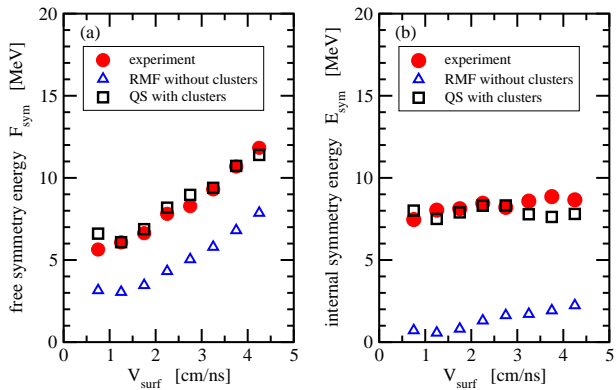}
\caption{\label{fig:01}%
(Color online) 
Free and internal symmetry energy as a function of the surface
velocity. Experimental results are compared with results of
theoretical calculations neglecting 
cluster formation (RMF) and including cluster formation (QS).}
\end{figure}

In Tab.~\ref{tab:01}, we also give results of the QS model \cite{Typel:2009sy} 
for the free and internal symmetry energies (columns 7 and 8) at given 
$T$ and $n$. 
In Fig.\ \ref{fig:01} (left panel) the experimentally obtained free 
symmetry energy is compared to the results of the RMF calculation without 
clusters and the QS model with clusters \cite{Typel:2009sy}.
There are large discrepancies between the measured values and the results of 
calculations in the mean-field approximation
when cluster formation is neglected.
On the other hand, the QS model results correspond nicely to the 
experimental data. 
In the right panel of Fig.~\ref{fig:01} we compare
the internal symmetry energy derived from the experimental data 
with the RMF and QS results. 
Again, it is clearly seen that the quasiparticle 
mean-field approach (RMF without clusters) 
disagrees strongly with the experimentally deduced symmetry 
energy while the QS approach 
gives a rather good agreement with the experimental data. 

\begin{figure}
\epsfig{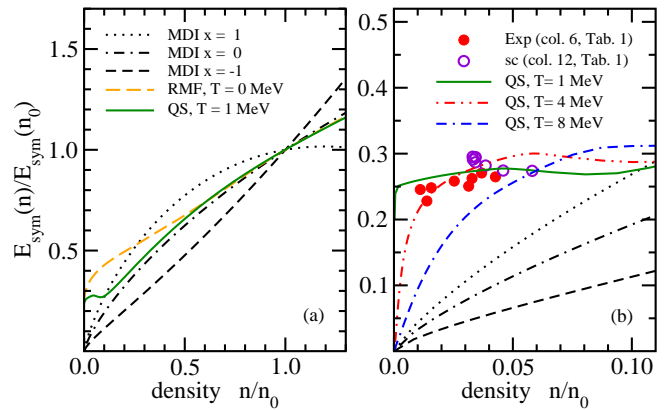}
\caption{\label{fig:02}
(Color online)
Comparisons of the scaled internal symmetry energy 
$E_{\rm sym}(n)/E_{\rm sym}(n_0)$ 
as a function of the scaled total density $n/n_0$ for different approaches 
and the experiment.
Left panel: 
The symmetry energies for the commonly used MDI parametrization of Chen et al. 
\cite{Che05} for $T=0$  
and different asy-stiffnesses, controlled by the parameter $x$ 
(dotted, dot-dashed and dashed (black) lines); 
for the QS model including light clusters
for temperature $T=1$~MeV  (solid (green) line), and for the
RMF model at $T=0$ including heavy clusters (long-dashed (orange) line).
Right panel:
The internal scaled symmetry energy in an expanded low-density region.
Shown are again the MDI curves and the QS results for $T=1,4,$ and 8~MeV
compared to the experimental data with the NSE entropy (solid circles) 
and the results of the self-consistent calculation (open circles) from 
Tab.~\ref{tab:01}}. 
\end{figure}

In Fig.~\ref{fig:02} we present results for different approaches to extracting 
the internal symmetry energy and compare with the experimental values.
In the left panel of the figure we show theoretical results for $T$ at or 
close to zero.
A widely used momentum-dependent parametrization of the symmetry energy (MDI) 
at temperature $T=0$~MeV was given in Refs.~\cite{Li:2008gp,Che05} and is 
shown 
for different assumed values of the stiffness parameter $x$.
For these parametrizations the symmetry energy vanishes in the low-density 
limit. We compare this to the QS result at $T=1$~MeV
(at lower $T$ crystallization or Bose condensation may occur as discussed 
above).
In this approach the symmetry energy is finite at low density.
The $T=1$~MeV 
curve will also approach zero at extremely low densities 
of the order of $10^{-5}~{\rm fm}^{-3}$ because 
the temperature is finite.
The RMF,~$T=0$ curve is discussed below. 
Also note that the underlying RMF model for the quasiparticle description 
with $n_0=0.149$~fm$^{-3}$, $E_{\rm sym}(n_0)=32.73$~MeV
gives a reasonable behavior at high density similar to the MDI,~$x$=0 
parametrization.
We thus see that our approach successfully interpolates between the clustering 
phenomena at low density and a realistic description around normal density.
 
In the right panel of Fig.~\ref{fig:02} we compare to the experimental 
results,
full (red) circles 
(Tab.~\ref{tab:01}, 
col. 6) 
in an expanded low-density region. 
Besides the MDI parametrization we show the QS results \cite{Typel:2009sy} for 
$T$=1, 4, and 8 MeV, which are in  the range of the temperatures in the 
experiment.  
The QS results including cluster formation agree well
with the experimental data points, as seen in detail in  Fig.~\ref{fig:01}.
We conclude that medium-dependent cluster formation has to be considered
in theoretical models
to obtain the low-density dependence of the symmetry energy
that is observed in experiments.

The temperatures and densities of columns 2 and 3  in Tab.~\ref{tab:01} will 
be  modified if medium effects on the light clusters are taken into account
\cite{Shlomo:2009}.
We have carried out a self-consistent determination of the temperatures
$T^{\rm sc}$ and densities $n^{\rm sc}$
taking into account the medium-dependent quasiparticle energies
as specified in Ref. \cite{Ropke:2008qk}
(columns 9 and 10 of Tab.\ \ref{tab:01}).
Compared to the Albergo method results \cite{Kowalski:2006ju}, the
temperatures $T^{\rm sc}$ are about 10 \% lower.
Significantly higher values are obtained for the inferred densities 
$n^{\rm sc}$ which are  more sensitive to the inclusion of medium effects.
We have also calculated the free and
internal symmetry energies corresponding to these self-consistent values  of 
$T^{\rm sc}$ and $n^{\rm sc}$ according to Ref.~\cite{Typel:2009sy} 
(columns 11 and 12 of Tab.~\ref{tab:01}).
These results are also shown in the right panel of Fig. 2 as open 
(purple) circles.
The resultant internal symmetry energies are 15 to 20 \% higher than the QS 
model values for $T$ and $n$ given in  columns 2 and 3  in 
Tab.~\ref{tab:01}.

We have restricted our present work to that region of the phase diagram 
where heavier clusters with $A > 4$ are not relevant. 
The generalization of the given approach to account for clusters of arbitrary 
size would lead to an improvement in the low-density, 
low-temperature region 
when nuclear statistical equilibrium is assumed.
Alternatively, one can introduce the formation of heavier nuclei 
in the presence of a nucleon and cluster gas, cf.\ 
Refs.~\cite{Lattimer:1991nc,Shen:1998gq}. 

The simplest approach to model the formation of heavy clusters is to perform 
inhomogeneous mean-field calculations in the Thomas-Fermi approximation
assuming spherical Wigner-Seitz cells. 
In Fig.~\ref{fig:02} (left panel) preliminary results for the zero-temperature 
symmetry energy of such a calculation 
is shown by the long-dashed line using the same RMF 
parametrization as for the QS approach
introduced above; for details see Ref.\ \cite{TypXXXX}.
The symmetry energy in this model approaches a finite value 
at zero density in contrast to the
behavior of the MDI parametrizations and conventional single-nucleon 
quasiparticle descriptions. 

In conclusion, we have shown that a quantum-statistical model of nuclear
matter, that includes the formation of clusters at densities below
nuclear saturation, describes quite well the low-density symmetry 
energy which was extracted from the analysis of heavy-ion collisions. 
Within such a theoretical approach
the composition and the thermodynamic quantities of nuclear matter
can be modeled in a 
large region of densities, temperatures and asymmetries that are
required,  e.g., in supernova simulations.

{\bf Acknowledgement}:
This research was supported by the DFG cluster of excellence ``Origin
and Structure of the Universe'', by CompStar, a Research Networking 
Programme of the European Science Foundation, by US Department of Energy
contract No.\ DE-AC02-06CH11357 (TK) and  grant No. DE-FG03-93ER40773 
(Texas A\& M) and by Robert A. Welch Foundation grant No. A0330 (JBN).
DB acknowledges support from the 
Polish Ministry for Research and Higher Education, grant 
No. N N 202 2318 37 and from the Russian Fund for 
Basic Research, grant No. 08-02-01003-a.

\end{document}